# REDUCING ENERGY CONSUMPTION OF NETWORK INFRASTRUCTURE USING SPECTRAL APPROACH

Khan Mohammad Habibullah[1], Eric Rondeau[2,3] and Jean-Philippe Georges [2,3]

[1] Université de Lorraine, Vandoeuvre-lès-Nancy, France.
[2] Université de Lorraine, CRAN, UMR 7039, Campus Sciences, BP 70239, Vandoeuvre-lès-Nancy Cedex, 54506, France
[3] CNRS, CRAN, UMR 7039, France.

## ABSTRACT

*The energy consumption by ICT (Information and Communication Technology) equipment is rapidly increasing which causes a significant economic and environmental problem. At present, the network infrastructure is becoming a large portion of the energy footprint in ICT. Thus, the concept of energy efficient or green networking has been introduced. Now one of the main concerns of network industry is to minimize energy consumption of network infrastructure because of the potential economic benefits, ethical responsibility, and its environmental impact. In this paper, the energy management strategies to reduce the energy consumed by network switches in LAN (Local Area Network) have been developed. According to the life-cycle assessment of network switches, the highest amount of energy is consumed during usage phase. The study considers bandwidth, link load and traffic matrices as input parameters which have the highest contribution to energy footprints of network switches during usage phase and energy consumption as output. Then with the objective of reducing energy usage of network infrastructure, the feasibility of putting Ethernet switches hibernate or sleep mode was investigated. After that, the network topology was reorganized using clustering method based on the spectral approach for putting network switches to hibernate or switched off mode considering the time and communications among them. Experimental results show the interest in this approach in terms of energy consumption.*

*.*

# INTRODUCTION

We are now living in the age of information and communication technology and the energy consumption, as well as carbon emission by computing and communications equipment is increasing rapidly. Mingay (2008) stated that, at present ICT is responsible for about 2% of global carbon emissions which is similar to the aviation industry. The enhancement of energy consumption has large negative economic and environmental impact. In recent years, we are struggling with climate change and in coming years it is the greatest challenge to tackle it. For that reason, energy consumption reduction is very important because it is directly related to greenhouse gas emission which is responsible for climate change. Recent research (Pamlin and Szomollanyi, 2006) suggests that, 15-30% volume of emission needs to be decreased in order to keep the global temperature increase below 2 degree Celsius.

The growth of internet users has been astonishing over the last few years. From 1990 to 2010, the number of Internet users increased from 3 million to 2 billion and by mid-2012, the user number had increased to 2.73 billion already. Regional growth rates between 2000 and 2012 have been exceeding 3 600% in Africa and 2 600% in the Middle East. As a result, the electricity demand of ICT is growing at much faster than overall electricity demand. Network-enabled device electricity demand is growing at a rate of 6% per year and total ICT energy demand reached 1560 TWh in 2013 (Iea.org, 2014). Because of this continuous growth of internet users, the spreading of broadband access and the increasing number of online services being offered by telecoms and Internet Service Providers, the energy efficiency issue has also become one of the highest priority objectives for wired networks and service infrastructures. These continuously rising trends in network energy consumption depend on new services being offered, as well as on data traffic volume increase (Nedevschi, et al., 2008). According to Zhang, et al. (2008), data traffic is increasing rapidly which follows Moore's law, by doubling every 18 months.

In recent years, some efficient steps have been taken to reduce the energy consumption of network infrastructure. This paper is a contribution to this mitigation in focusing only to reduce the energy consumption of network switches in a LAN (Local Area Network). Mahadevan, et al., (2010) stated that, a network switch consumes energy during its manufacturing phase, usage phase and dismantling phase with the highest amount of energy consumed by network switch during usage phase.

The concern of this research work is only during the usage phase. The objective is to reduce the energy consumption that simple network architecture builds up with Ethernet Switches and end devices. It is described in (Iea.org, 2014), up to 80% of energy consumption is used for some devices just to maintain the network connection and more than 600 TWh of electricity was consumed on 2013 by such devices. It is possible to reduce up to 65% of energy consumption of such devices using the best available technologies and strategies. Hossain, et al., (2015) and Gunaratne, et al., (2008) found bandwidth and number of connection are mostly responsible for energy consumption of Ethernet switch in usage phase. Krommenacker, et al., (2001) and Rondeau, et al., (2001) showed proper network architecture designs and cabling plan using clustering approach increases the efficiency of the network. An efficient network design reduces traffic propagation delay, jitter, and packet loss which also reduces energy consumption. Therefore, in this paper, optimization, reorganization and clustering approach of network architecture have been proposed by analyzing bandwidth and the traffic load in point to point communication. The feasibility to hibernate or switch off devices which only remain alive to maintain network connection was also investigated, with the objective of reducing energy consumption.

The experiment was done with cisco switches and strategies have been applied to optimize and reorganize the network topology. In the experiment, the power consumption was used to measure energy usage of network architecture. In the first phase of experiment, we designed network topology consists of connected Ethernet switches which support energywise and desktop computers were connected to each switch. Then we measured the power consumption of all switches using different bandwidth with same traffic load according to traffic matrix. In the second phase, we reorganized and optimized our network architecture using clustering method based on spectral approach, with the objective to reorganize network cabling in order to be able to sleep or hibernate unused components of network infrastructures for a certain amount of time during low traffic periods.

Hibernation mode decreases a large amount of power consumption compared to the operation mode of an Ethernet switch and switched off mode does not consume any power. In the third phase, we investigated the feasibility of putting various components connected with Ethernet switches or the switch itself to sleep or hibernate in considering the changes of communication activities. After applying the spectral approach to different cases of communication with the same network topology, it was possible to hibernate or switch off the network switch which saved a significant amount of energy.

## RELATED WORK

Numerous studies have explored both in the wired and wireless network to analyze the energy consumption pattern of network devices and to reduce energy consumption using different strategies.

**Related Work in Network Energy Consumption Evaluation**

Christensen, et al., (2004) have studied the lifecycle energy use of network devices and explained how network devices can impact on environment pollution. The lifecycle energy use of network devices has been studied by Nordman (2008), but mostly in the context of home and office environments. Rivoire, et al., (2007) explained from a device manufacturer's point of view that, network devices such as routers and switches are power proportional which means they consume energy proportional to their usage. Gupta, et al., (2004) showed how different parameters affect the energy consumption of network switch. After that, they defined a model which shows the relationship between parameters related to the Ethernet switch and energy consumption. Fithritama, et al., (2015) proposed a method based on fuzzy logic to identify the relationships between network parameters and their effect on power consumption when deploying new network equipment. Their proposed method is also applicable to control the desired level of energy consumption by tuning the network parameters. In the research by Reviriego, et al., (2012), the energy consumption of small Energy Efficient Ethernet (EEE) switches is analyzed in several experiments. Based on the experiment result, the authors proposed a model for the energy consumption of Energy Efficient Ethernet switch.

**Related Work in Network Energy Consumption Optimization**

Gupta, et al., (2004) explored the feasibility of power management schemes at network switches in the LAN. They examined the possibility to put various components on LAN switches to sleep for reducing energy consumption. Experiments to evaluate energy management studies for network switches were performed by Mahadevan, et al., (2010). The authors found, energy usage in the operational stage is dominating and they parametrically examine various energy management techniques to reduce the operational energy footprint of network switches. Nedevschi, et al., (2008) presented the design and evaluation of two forms of power management schemes to reduce the energy consumption of network. They have shown that simple schemes for sleeping or rate adaptation have significant energy savings without noticeably increased packet loss and latency.

**Related Work Based on Clustering Approach**

Consideration of information flows is very important to design network architecture. The objective of network designer is to confine the strong co-operation and communication with sub-network to avoid flooding and overloading the whole network. In order to minimize overloading, intra-group communication should be maximum and inter-group communication should be minimum. To achieve this network scalability objective, clustering method has been widely pursued by the research community. In this research work, clustering algorithm has been used to reduce communication; isolate groups for hibernating or switched off part of the network.

Many clustering algorithms have been proposed (Abbasi, et al., 2007; Amis, et al., 2007; Baker, 1981; Bandyopadhyay, 2003; Basagni, 1999; Lin, 1997; Mellier, 2006; Chiasserini, et al., 2002) to fulfill various objectives based on wireless sensor network but none of these algorithms aim to minimize energy consumption in network architecture. Most of these algorithms are heuristic in nature and the criteria for cluster selection and node grouping are intra and inter-cluster connectivity if the application is sensitive to data latency and the length of data routing paths. The objective of these algorithms is to generate the minimum number of clusters. White, et al., (2005) proposed the spectral algorithm to find communities in a graph and showed that spectral algorithm is effective and efficient at finding both good clustering and the appropriate number of clusters from a variety of real-world graph data sets. They observed spectral algorithm is faster for large sparse graphs. Von Luxburg (2007) introduced and presented the most common spectral algorithm from scratch by different approaches. He also discussed the advantages and disadvantages of different spectral clustering algorithms, different graph Laplacians and their basic properties. Krommenacker, et al., (2001) defined some criteria to reorganize the network architecture and explored algorithms, especially the spectral algorithm to define cabling plan of switched network for real-time applications. The spectral algorithm to design the cabling plan for industrial Ethernet architecture has been used by Rondeau, et al., (2001), which can reduce handling delays of messages inside the time cycle of applications.

## SYSTEM ARCHITECTURE AND CRITERIA TO REORGANIZE NETWORK SWITCHES

This section describes the general network architecture that has been used in the research and the criteria to reorganize and design efficient architecture for switched Ethernet Networks. Figure 1 shows the network architecture and the cabling plan used during the experiment. Three Cisco

Ethernet Switch 2960-X (S1, S2, and S3) were used during the experiment and three desktop PCs (Personal Computer) were connected to each switch.

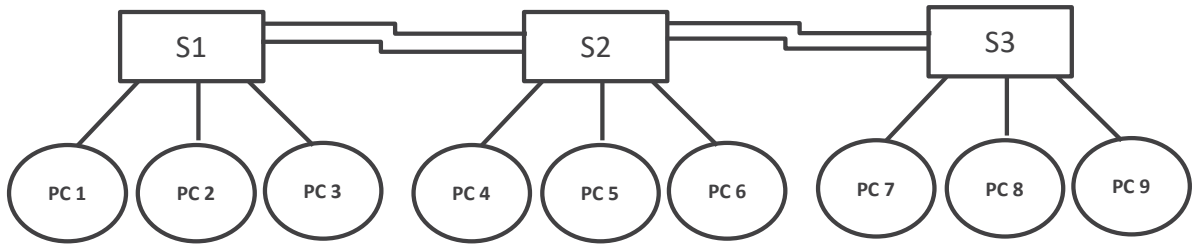

*Figure 1: Cabling plan and Network architecture.*

In the section below we have described the criteria to reorganize network switches in different actions.

**Action 1:**
It illustrates that if PC 1 has frequent communication with large data size with PC 7, then it is better to connect either PC1 in S3 or PC7 in S1. Thus, it is very important to consider information flows during designing a network. In this case, the goal is to maintain communication among devices by avoiding overloading the whole network which will increase network efficiency and reduce energy consumption.

**Action 2:**
The objective of action 2 is to gather the PCs strongly communicating among them, with the idea to group PCs without communication in other clusters in order to be able to hibernate or switch off these clusters.

**Action 3:**
If the capacity of each port of all switches is 1 Gbps and total traffic load for the communication between S1 and S2 or S2 and S3 are more than 1Gbps, then more than one trunk port needs to be connected to support the communication. In this situation, if inter-switch communication can be reduced, it is possible to eliminate redundant trunk between two switches. Thus reducing the traffic between switches allows reducing bandwidth and as a result, reduction of energy consumption.

Therefore, in general, the criteria to reorganize the communication among switches are to minimize inter-group dialogues in order to maximize the intra-group exchanges or to create groups without communication. Some other criteria should also take into account such as the number of ports and capacity of each port with the objective to use the smaller number of switches with a higher number of connected ports.

## METHODOLOGY

The spectral algorithm is used to cluster the end device according to the usage of switch ports and to ensure the minimal interaction between two clusters. As an initial step of the approach, a graph $G_d$ has been introduced and translated that into traffic matrix according to the communication among end devices.

This section describes clustering method based on the spectral approach which has been used to reorganize and optimize network topology.

**Spectral Approach**

The spectral algorithm is one of the most successful heuristics for partitioning graphs and matrices. This approach is used to solve scientific numerical problems such as solving parse linear systems explained by Pothen, et al., (1992), partitioning for domain decomposition (Chan and Smith, 1994), Ethernet architecture segmentation (Rondeau, et al., 2001) and so on. In this research spectral approach has been used to

- Adjust the size of the clusters in regard to the capacity of the switches in terms of port number.
- Reorganize and minimize the cabling of the switch by breaking graphs into sub graphs.
- Investigate the feasibility of putting unused switches to hibernate or switched off according to usage.

Given an undirected weighted graph $G = (V, E)$, where G is a division of its vertices $v$ into two disjoint subsets, $V_1$ and $V_2$. Let $E(V_1, V_2)$ is the set of edges (links for communication) $e_{ij}$ with one endpoint in $V_1$ and other in $V_2$. The cut size of the partition is the sum of the edge weights of of E :

$$c = \sum_{eij \in E} w(eij)$$

Where $v_i \in V_1$ and $v_j \in V_2$

Here the objective is to find a disjoint subset of $V, P = \{v_i\}$ which minimizes $c$.
The adjacency matrix $A(G_d)$, of the graph G is $n \times n$ matrix where (ij)th entry is the weight of edge $e_{ij}$ : $w_{ij}$ and 0 otherwise. For adjacency matrix, the diagonal entries are always 0.

As the spectral algorithm considers only undirected graph, thus it needs to work with
$A(G) = A(G_d) + A(G_d)^t$

**Algorithm Description**

Let D (G) is an undirected graph converted into $n \times n$ diagonal matrix such as
$$d_{ii} = \sum_j w(eij)$$
The Laplacian matrix $L(G)$ of the graph G is $L(G) = D(G) - A(G)$.
According to (Donath, et al., 1973; Fiedler, 1973, 1975a, 1975b) a good cut size is the second smallest eigenvalue λ2 and its associated eigenvector $\vec{u} = (u_1, u_2, \ldots u_n)$ (Fiedler vector).
Now, sorting the vertices on the incremental way according to the values of the components of Fiedler vector gives the reorganization of the vertices.

The objective of the partitioning is to use exactly d devices (in this case network switches) and to minimize the communication through the switches if there are two levels of switches. A splitting value s should be found for spectral partitioning which divides the vertices of G into $V_1$ such that $u_i > s$, and $V_2$ such that $u_i < s$ and that is called a Fiedler cut. The choices for the splitting values s are:

- Bisection cut, where s is the median of $(u_1, u_2, \ldots u_n)$.

- Sign cut, where s is equal to 0.
- Ratio cut, where s is the value that gives the best cut ratio denoted $\emptyset$ with $\emptyset(V_1, V_2) = |E(V_1, V_2)|/\min(|V_1|, |V_2|)$
- Gap cut, where s is the value in the largest gap in the sorted list of the Fielder vector components.

To partition a graph into the power of two numbers of groups, Recursive Spectral Bisection (RSB) algorithm explained by Simon (1991) is applied. In this paper, optimized recursive bisection algorithm has been used for partitioning the graph.

**Optimized Recursive Spectral Bisection Algorithm (RSB in $[d/2]$ )**

Simon (1991) proposed improvement on RSB algorithm for designing a switched Ethernet network cabling plan as follows:
1. Compute the Fiedler vector for the graph.
2. Sort the vertices according to the values of the components of the Fielder vector.
3. Calculate d for the group to be partitioned.
4. Assign $n[d/2]$ first vertices to one sub-group and the others to the second one.
   Cut of value ($n[d/2]$) = number of ports * [number of devices/2].
5. Apply steps 1-4 recursively to each sub-group until the size of each sub-group becomes$\leq$ n.

## RESEARCH METHOD AND EXPERIMENT DETAILS

**Parameter Selection**

Parameter selection is one of the most important issues for power management of Ethernet switch. Bandwidth and the number of connections are mostly responsible for energy consumption of Ethernet switch (Hossain, et al., 2015) and the packet size has no impact on energy consumption of the switch (Mahadevan, et al., 2009).So in our experiment bandwidth, the traffic load on each switch port and traffic matrix according to the communication among end devices was used as the input parameter and power consumption of switch was measured as the output parameter. Though traffic load has very little impact on energy usage, but it represents the service that the network must offer (QoS). In the experiment traffic load was used to maintain the frequency of communication among the end devices and to draw traffic matrix depending on that communication.

**Detailed Experiment**

The experiment was done according to the figure 1. Powerspy2 sensor was used to measure the total energy consumption of three switches which can send real-time energy consumption data via Bluetooth (Hossain, et al., 2015 and PowerSpy2 user manual, 2015). Ostinato traffic generator was used to generate traffic into the switch ports (Ostinato Network Traffic generator and Analyzar User Guide). Three sets of experiment were done and each set contains two experiments. During each set of experiment, for all switch ports, the same bandwidth was used which were 100Mbps and 1Gbps. Depending on the traffic load and frequency of communication we have designed $9 \times 9$ traffic matrix and measured energy consumption of the network topology. According to the frequency of communication among end devices, two types of traffic loads has been used which are defined as large transmission and small transmission which are shown in traffic matrix as "10" and "1" in respectively. More information about traffic loads and frequency of communication is shown in Table 1. Each experiment was run and monitored for 15 minutes.

|  | For large transmission (10) | For small transmission (1) |
|---|---|---|
| **Frame size (Byte) :** | 1125 | 1125 |
| **Packet per second:** | 10000 | 1000 |
| **Total size of transmission per second (Mbps)** | 90 | 9 |

**Table 1: Traffic load and frequency of communication.**

## RESULTS AND DISCUSSIONS

This section presents the obtained result during the test phase. In traffic matrix, the red marked PCs are connected with S1, blue marked PCs are connected with S2 and green marked PCs are connected with S3.

**Case Study**

For conducting the experiment, communication among the end devices during day time or working hours and night time or nonworking hours was considered. The working hour was defined as $T_1 = 16$ hours and the nonworking hour was defined as $T_2 = 8$ hours. The communication among the end devices is more than nonworking hours which have been shown by traffic matrix1 and matrix2 accordingly in Figure 7. Both of the matrices in Figure 7 are unoptimized.

The architecture is composed of 9 PCs connected to 3 Ethernet switches (respectively S1, S2, and S3) with the capacity of five ports each switch as explained in figure 1. Figure 7 shows that PCs in red (1, 2 and 3) are connected to switch S1, PCs in blue (4, 5 and 6) to switch S2 and PCs in green (7, 8 and 9) to switch S3.

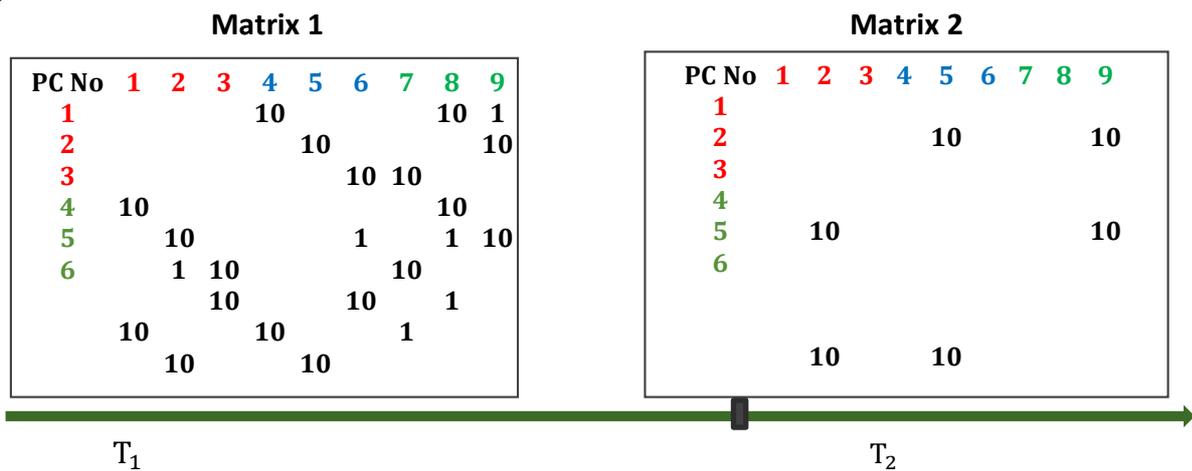

**Figure 7: Traffic matrix according to the communication among end devices during working and non-working hours.**

In this case, most of the frequent communications are intra-switch communication and in optimized organization, most of the frequent communications are inter-switch communication among end devices showed in Figure 7. The total amount of traffic load is same for both matrices.

The yearly power consumption for Figure 7 can be calculated by the formula below:
$$P_{total} = (T_1 P_{matrix1} + T_2 P_{matrix2}) * 365$$
Here $P_{total}$ is the yearly power consumption, $T_1 P_{matrix1}$ is daily power consumption for Matrix 1 and $T_2 P_{matrix2}$ is daily power consumption for Matrix 2.

Table 2 and Table 3 shows the experimental results of power consumed by the network architecture showed in Figure 1 for 100 Mbps and 1Gbps bandwidth. For all measurements, bandwidth for both trunk ports and the ports connected to end devices was same and two trunk ports were connected between each switch to support communications.

| Time | Working hours ($T_1$) | Non-working hours ($T_2$) | Total |
|---|---|---|---|
| Power consumption (KWh/Year) | 598.6 | 299.6 | 898.2 |

Table 2: Power consumption for 100 Mbps bandwidth in the unoptimized organization.

| Time | Working hours ($T_1$) | Non-working hours ($T_2$) | Total |
|---|---|---|---|
| Power consumption (KWh/Year) | 620.8 | 308.9 | 929.7 |

Table 3: Power consumption for 1 Gbps bandwidth in the unoptimized organization.

There is no notable change in power consumption for traffic load, but there is a noticeable change in power consumption when there is a change in bandwidth. The power consumption increases 0.2 to 0.4 Watt per port if the link speed is changed from 100 Mbps to 1Gbps respectively. Reducing bandwidth from 1 Gbps to 100 Mbps could save yearly 33.3 KWh in the unoptimized organization where 5 ports in each switch were active. Another noticeable observation is, an extra trunk link was needed in working hours for the unoptimized organization to support communication and to avoid packet loss for traffic overload between S1 and S2.

For clustering and optimizing the communication and cabling plan for the whole day, the weights of Matrix 1 and Matrix 2 have been calculated considering the time $T_1$ and $T_2$ respectively and the spectral algorithm was implemented on that matrix. It has been assumed that Matrix 1 and Matrix 2 are derived from connected graph G1 and G2 respectively. The weight of G1 and G2 is defined by C which is shown in Figure 8.

| PC No | 1 | 2 | 3 | 4 | 5 | 6 | 7 | 8 | 9 |
|---|---|---|---|---|---|---|---|---|---|
| 1 |   |   |   |   |   | 160 |   | 160 | 16 |
| 2 |   |   |   |   | 240 |   |   |   | 240 |
| 3 |   |   |   | 160 | 160 |   |   |   |   |
| 4 | 160 |   |   |   |   |   |   | 160 |   |
| 5 |   | 240 |   |   |   | 16 |   | 16 | 240 |
| 6 |   | 16 | 160 |   |   |   | 160 |   |   |
| 7 |   |   | 160 |   |   | 160 |   | 16 |   |
| 8 | 160 |   |   | 160 |   |   | 16 |   |   |
| 9 |   | 240 |   | 240 |   |   |   |   |   |

$$C = T_1G1 + T_2G2$$

| PC No | 1 | 2 | 3 | 4 | 5 | 6 | 7 | 8 | 9 |
|---|---|---|---|---|---|---|---|---|---|
| 1 |   |   |   |   | 320 |   |   | 320 | 16 |
| 2 |   |   |   |   | 480 | 16 |   |   | 480 |
| 3 |   |   |   | 320 | 320 |   |   |   |   |
| 4 | 320 |   |   |   |   |   |   | 320 |   |
| 5 |   | 480 |   |   |   | 16 |   | 16 | 480 |
| 6 |   | 16 | 320 |   |   | 16 | 320 |   |   |
| 7 |   |   | 320 |   |   | 320 |   | 32 |   |
| 8 | 320 |   |   | 320 | 16 |   | 32 |   |   |
| 9 |   | 16 | 480 |   |   | 480 |   |   |   |

A(C)

**Figure 8: Combined matrix and undirected adjacency matrix for the communication.**

As the spectral algorithm works with only undirected graph, so A (C) = C+C' has been calculated shown in Figure 8 and from there Laplacian matrix L(C) has been calculated which is shown in Figure 9.

| $V_{FIEDLER}$ | Sorted Vertices |
|---|---|
| 0.45919 | 3 |
| -0.12275 | 6 |
| -0.34744 | 7 |
| 0.46906 | 2 |
| -0.11657 | 5 |
| -0.33986 | 9 |
| -0.32144 | 8 |
| 0.43358 | 1 |
| -0.11378 | 4 |

L(C):

|   | 5 | 6 | 7 | 8 | 9 |
|---|---|---|---|---|---|
|   |   |   |   | −320 | −16 |
|   | −480 | −16 |   |   | −480 |
|   |   | −320 | −320 |   |   |
|   |   |   |   | −320 |   |
|   | 992 | −16 |   | −16 | −480 |
|   | −16 | 672 | −320 |   |   |
|   |   | −320 | 672 | −32 |   |
|   | −16 |   | −32 | 688 |   |
|   | −480 |   |   |   | 976 |

**Figure 9: Laplacian matrix for graph C**

**Figure 10: Fiedler values and sorted vertices for the Graph G.**

After computing the Fielder vector for the graph C and sorting the vertices according to the values of the components of the Fielder vector, sorted vertices showed in Figure 10 were obtained.

Here nine end devices are connected to three network switches with the capacity of five ports where two ports are connected as trunk. The sorted vertices from Figure 10 have been used as end devices. After applying improvement on RSB algorithm (Rondeau, et al., 2001), the obtained result of Figure 11 partitioned the end devices into three parts and the group of three end devices is connected to each switch.

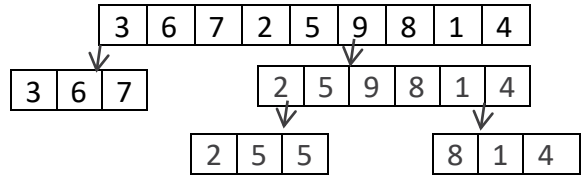

*Figure 11: Partitioning of End Devices by Using RSB in $n[d/2]$*

The optimized adjacency matrix was drawn based on working hours and nonworking hours according to the serialization found in Figure 12.

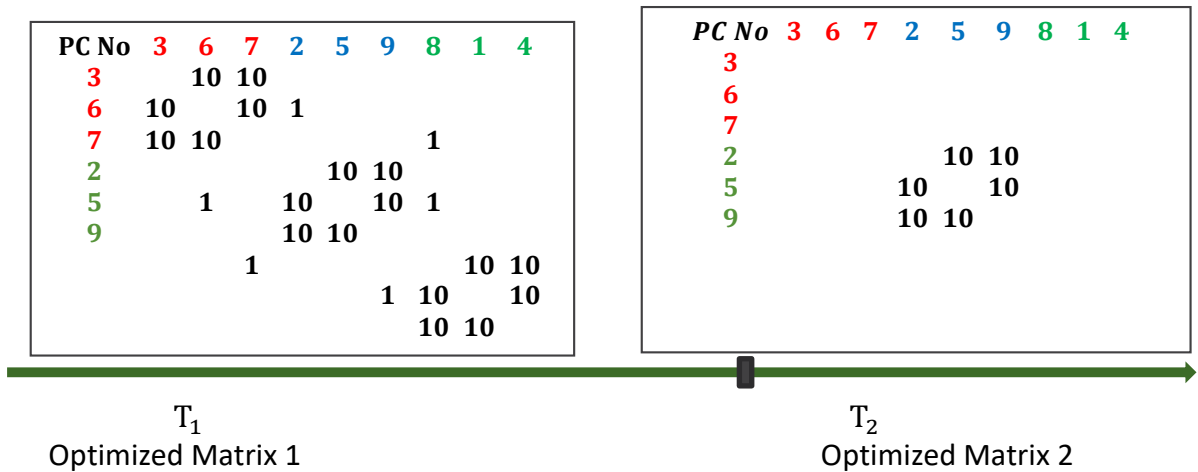

*Figure 12: Optimized Matrix 1 and Matrix 2.*

The power consumption by the optimized solution is measured which is showed in Table 4.

**Table 4: Power consumption for working hour and nonworking hours for optimized solution.**

| Power Consumption at Working hours (T1) by Optimized Organization | | |
|---|---|---|
| Bandwidth | 100 Mbps | 1 Gbps |
| Energy consumption (KWh/Year) | 596.3 | 613.8 |

| Power Consumption at Nonworking hours (T2) by Optimized Organization (KWh/year) | | | |
|---|---|---|---|
| Bandwidth | With Trafic | Hibernate | Switched off |
| 100 Mbps | 298.7 | 218.1 | 99.9 |
| 1 Gbps | 306.9 | 219.6 | 100.7 |

**Table 5: Total yearly power consumption for the optimized solution for different bandwidth.**

| | Power consumption for 100Mbps bandwidth (KWh/year) | Power consumption for 1Gbps bandwidth (KWh/year) |
|---|---|---|
| **Always active** | 895.0 | 920.7 |
| **With hibernation** | 814.4 | 833.4 |
| **With Switched off** | 596.1 | 613.8 |

Table 5 shows the yearly total power consumption of optimized solution using the spectral algorithm. The result shows that optimized solution consumes less energy than unoptimized architecture. It was also possible to reduce extra trunk port between S1 and S2 during working hours because spectral algorithm produced clusters of PCs which have strong communications among those which reduce inter-switch communication showed in optimized Matrix 1. After applying spectral clustering on end devices and serialization of network switches, it was possible to hibernate or switch off two network switches (S1 and S3) during nonworking hours because it created two clusters of PCs without communication which are connected with S1 and S3.

**Table 6: Power saving by using optimized solution**

|  | Power saving for 100Mbps bandwidth (KWh/year) | Power Saving for 1Gbps bandwidth (KWh/year) |
|---|---|---|
| **Always active** | 3.2 | 9.0 |
| **With hibernation** | 83.8 | 96.3 |
| **With Switched off** | 202.1 | 315.9 |

Table 6 shows the power savings after using the spectral algorithm for different bandwidth. For 100 Mbps bandwidth, the optimized solution could save a yearly minimum of 3.2 KWh when no hibernate or switched off feature activated to maximum 202.1 KWh with switched off feature activated during nonworking hours. For 1Gbps bandwidth, the optimized solution could save 9.0 KWh when no hibernate or switched off feature activated to maximum 315.9 KWh for switched off feature activated during nonworking hours.

For hibernation or switched off Ethernet switch, it should be considered the Quality of Service. The hibernation mode consumes about 20 watts and switched off mode consumes nothing. For the Cisco switch model 2960-X, it takes 260 seconds to get ready after wake up from hibernation and 290 seconds from switched off mode. The time difference to be fully functional from hibernation and switched off mode is 30 seconds which can be critical for few cases. But for normal office works or Local Area Network, it doesn't have that much impact. Frequent on-off can put extra load which can reduce lifetime and some devices are vulnerable to this situation. On the other hand, always keeping on a device can also reduce lifetime, therefore switched off can be a good choice too. Hibernation and wake up time can be scheduled by command line which is not possible for switched off mode but by using rack power distribution units (PDUs) it is also possible to switch off and on automatically. Therefore, the choice of hibernation or switched off Ethernet switch can be according to the necessity of the user.

The spectral approach is also applicable for optimizing bigger network architecture. So by using clustering method based on spectral approach, it is possible to save a large amount of energy according to traffic loads and pattern for the bigger picture and more complex architecture. The Ethernet switch is only a part of network architecture and to put efficient and greater impact, other network components like router, WIFI hotspot etc. should be considered. It is also possible to apply the spectral algorithm for clustering those devices according to traffic and optimize the whole architecture. Moreover, the experiment was done with one type of Ethernet switch, so the result may vary for different type of switches. Rondeau, et al., (2015) mentioned the correlation between the power consumption and carbon emission. Therefore, by optimizing network architecture using the spectral algorithm, it is also possible to reduce carbon footprint caused by network equipment. As a big picture, with proper design and development of network architecture, it is possible to reduce and control global energy usage and carbon footprint caused by network devices.

## CONCLUSION

The enhancement of energy consumption by network infrastructures has large negative impact on sustainability which should be controlled. This paper presents a novel way of energy efficient network design and management system by using the clustering approach based on spectral algorithm to reduce energy consumption of network infrastructure. By using this approach, it is possible to hibernate or switch off part of network during low traffic hours in order to save energy. Experimental result shows that significant amount of energy can be saved by reorganizing the network switches and clustering the end devices using the spectral algorithm. For future work, this approach can be applied to the bigger network architecture and big enterprise which has the potential to save a large amount of energy usage. Similar experiments can be done for other network devices such as routers to reduce more energy consumption.

**Acknowledgement:** This work is part of the Erasmus Mundus Master program in Pervasive Computing and Communication for Sustainable Development (PERCCOM) of the European Union (www.perccom.eu). The authors thank all the partner institutions, sponsors and researchers of the PERCCOM program (Klimova, et al., 2016 ).